\documentclass[11pt,twoside]{article}

\usepackage{asp2006}
\usepackage{epsf}
\usepackage{psfig}
\usepackage{lscape}
\def\ltsima{$\; \buildrel < \over \sim \;$}
\def\simlt{\lower.5ex\hbox{\ltsima}}
\def\gtsima{$\; \buildrel > \over \sim \;$}
\def\simgt{\lower.5ex\hbox{\gtsima}}

\newcommand\mbh{{\,{\rm M}_{\rm bh}}}
\def\msun{{\,{\rm M}_\odot}}

\begin{document}
\title{Feedback from Nuclear Star Clusters (NCs) and SMBHs}   
\author{Sergei Nayakshin}   
\affil{Department of Physics \& Astronomy,  University of Leicester, Leicester, LE1 7RH, UK}    

\begin{abstract} 
The observed super-massive black hole (SMBH) mass -- galaxy velocity
dispersion ($M_{\rm cmo} - \sigma$) correlation, and the similar correlation
for nuclear star clusters, may be established when winds/outflows from the CMO
(``central massive object'') drive gas out of the potential wells of classical
bulges. Timescales of growth for these objects may explain why smaller bulges
appear to host preferentially NCs while larger ones contain only SMBHs.

Despite much recent progress, feedback processes in bulge/galaxy formation are
far from being understood. Our numerical simulations show that that
understanding how the CMO feeds is as important a piece of the puzzle as
understanding how its feedback affects its host galaxy. 
\end{abstract}


\section{Analytical arguments for $M_{\rm cmo} - \sigma$ relations}  

It is believed that the centres of most galaxies contain SMBHs whose mass
$M_{\rm bh}$ correlates with the velocity dispersion $\sigma$ of the host
galaxy \citep{Ferrarese00,Gebhardt00,Tremaine02}. Similarly, there is a
correlation between $\mbh$ and the mass of the bulge, $M_{\rm bulge}$ for
large SMBH masses \citep{Magorrian98,Haering04,GultekinEtal09}.  Observations
also suggest that the masses of NCs ($10^5 \msun \simlt M_{\rm NC} \simlt 10^8
\msun$) correlate with the properties of their host dwarf ellipticals
\citep{FerrareseEtal06,Wehner06} in a manner that is analogous to the one
between SMBHs and their host ellipticals.

These relations can be explained in a similar way if the growth of host
galaxies and their central SMBHs or NSCs are linked by momentum feedback.  In
the model of \cite{King03,King05}, the SMBH luminosity is assumed to be
limited by the Eddington value. Radiation pressure drives a wind, the momentum
outflow rate of which is
\begin{equation}
\dot P_{\rm SMBH} \approx {L_{\rm Edd}\over c} = \frac{ 4 \pi G M_{\rm BH}}{\kappa}\;;
\label{pismbh}
\end{equation}
\noindent here $\kappa$ is the electron scattering opacity and $M_{\rm BH}$ is
the SMBH mass. Because the cooling time of the shocked gas is short on scales
appropriate for observed bulges, the bulk energy of the outflow is thermalised
and quickly radiated away. It is then only the momentum push (equation
\ref{pismbh}) of the outflow on the ambient gas that is important since it is
this that produces the outward force on the gas. The weight of the gas is
$W(R) = GM(R)[M_{\rm total}(R)]/ R^2$, where $M_{\rm gas}(R)$ is the enclosed
gas mass at radius $R$ and $M_{\rm total}(R)$ is the total enclosed mass
including dark matter. For an isothermal potential, $M_{\rm gas}(R)$ and
$M_{\rm total}(R)$ are proportional to $R$, so the result is
\begin{equation}
W = {4f_g\sigma^4\over G}\;.
\label{w}
\end{equation}
Here $f_g$ is the baryonic fraction and $\sigma^2 = GM_{\rm total}(R)/2R$ is
the velocity dispersion in the bulge.
By requiring that momentum output produced by the black hole just balances
the weight of the gas, it follows that
\begin{equation}
M_{\sigma} = {f_g\kappa\over \pi G^2}\sigma^4,
\label{msigma}
\end{equation}
\noindent which is consistent with the observed the $M_{\rm BH}$--$\sigma$ relation.

\cite{McLaughlinEtal06} proposed that the observed $M_{\rm NC} -\sigma$
relation for dwarf elliptical galaxies follows naturally from an extension of
the above argument (\cite{King03,King05}) to the outflows from young star
clusters containing massive stars. These individual stars are also
Eddington--limited, and produce outflows with momentum outflow rate $\sim
L_{\rm Edd}/c$ where $L_{\rm Edd}$ is calculated from the star's mass.  Young
star clusters with normal IMFs produce momentum outflow rate
\begin{equation}
\dot\Pi_{\rm NC} \approx \lambda{L_{\rm Edd}\over c}
\label{pinc}
\end{equation}
where $\lambda \approx 0.05$ and $L_{\rm Edd}$ is now formally the
Eddington value corresponding to the total cluster mass. To produce
the same amount of momentum feedback, a young star cluster must
therefore be $1/\lambda$ times more massive than a SMBH radiating at
the Eddington limit, and hence:
\begin{equation}
M_{\rm NC} = {f_g\kappa\over \lambda\pi G^2}\sigma^4.
\label{msignc}
\end{equation}
Strikingly, $1/\lambda$ is quite close to the offset in mass between the
$M_{\rm BH}$--$\sigma$ and $M_{\rm NC}$--$\sigma$ relations. 

\cite{NayakshinEtal09b} noted that timescales are important in this problem as
well as energetics. SMBH growth is limited by the Eddington accretion rate,
$\dot M_{\rm Edd} = L_{\rm Edd}/(\epsilon c^2)$, where $\epsilon \sim
0.1$ is the radiative efficiency of accretion.  SMBH masses can grow
no faster than $\exp(t/t_{\rm Salp})$, where
\begin{equation}
t_{\rm Salp} = \frac{M_{\rm BH}}{\dot M_{\rm
  Edd}} = \frac{\kappa \epsilon c}{4\pi G} = 4.5\times
10^7\epsilon_{0.1}~{\rm yr}
\label{tsalp}
\end{equation}
is the Salpeter time, with $\epsilon_{0.1} = \epsilon/0.1$. Star formation can
occur on the free--fall or dynamical timescale $t_{\rm dyn}$ of the system,
which is less than a million years for many observed young star clusters.
\cite{NayakshinEtal09b} obtained for the dynamical time as a function of
velocity dispersion:
\begin{equation}
t_{\rm dyn} =
17\left(\frac{\sigma}{150\,\mbox{km\,s}^{-1}}\right)^{2.06} \mbox{Myr}
\end{equation}
A simple theory for the observed bimodality of NC and SMBHs was offered. In
galaxies with small velocity dispersions ($\sigma \simlt 150$ km s${-1}$),
star formation occurs more rapidly than SMBH growth. NCs reach their maximum
mass and drive the gas away before the hole can grow. The opposite occurs in
galaxies with larger velocity dispersions, where SMBH are able to grow quickly
enough to reach their maximum ($M_{\rm BH}$--$\sigma$) mass. Low--dispersion
bulges thus have underweight SMBHs, while high--dispersion bulges do not have
nuclear star clusters.

\section{Numerical models}

One would clearly like to test these attractively simple explanations with
careful numerical models of SMBH/NCs growth and feedback on the host. We have
recently implemented a novel feedback and gas accretion model for SMBH
\citep{NayakshinEtal09a,NP10}. SMBH wind outflows are modelled with
collisionless ``momentum particles'', whereas accretion model follows the
``sink particle'' approach from star formation. In contrast to the Bondi-Hoyle
models, our model obeys the angular momentum conservation law, e.g., accretes
only low angular momentum gas, allowing us to follow accretion
\citep{HobbsEtal2010} and feedback in cases when discs form.

Simulating accretion on the SMBH and its feedback on gaseous shells in a
static isothermal potential, \cite{NP10} shows that the spherically symmetric
initial conditions reproduce the \cite{King03,King05} results
excellently. However, when rotation or SMBH flow collimation are added to the
initial conditions, the results get considerably more complicated, to the
point that it is not clear how any robust $M_{\rm cmo} - \sigma$ relation can
be established. We emphasise that this is a completely general problem common
to any {\em realistic} feedback model, as rotation or other symmetry breaking
changes the one-to-one relation between accretion and feedback of the simple
spherical models.

The following simulation of a collimated SMBH outflow misaligned with the disc
demonstrates the general problem here. The simulation starts with a spherical
rotating shell of gas far from the SMBH.  For simplicity, the SMBH momentum
outflow rate is fixed at the Eddington rate as described in the previous
section. The outflow is constrained to the conical surface with opening angle
$\theta = 45^\circ$ around the direction of the axis of symmetry of the
outflow. The latter is inclined by angle $\pi/4$ from the $z$-axis defined by
the direction of the angular momentum vector of the shell. For convenience of
presentation, we choose the angular momentum vector of the shell and the
outflow symmetry axis to lie in the $z-y$ plane.

\begin{figure}[!ht]
\centerline{\psfig{file=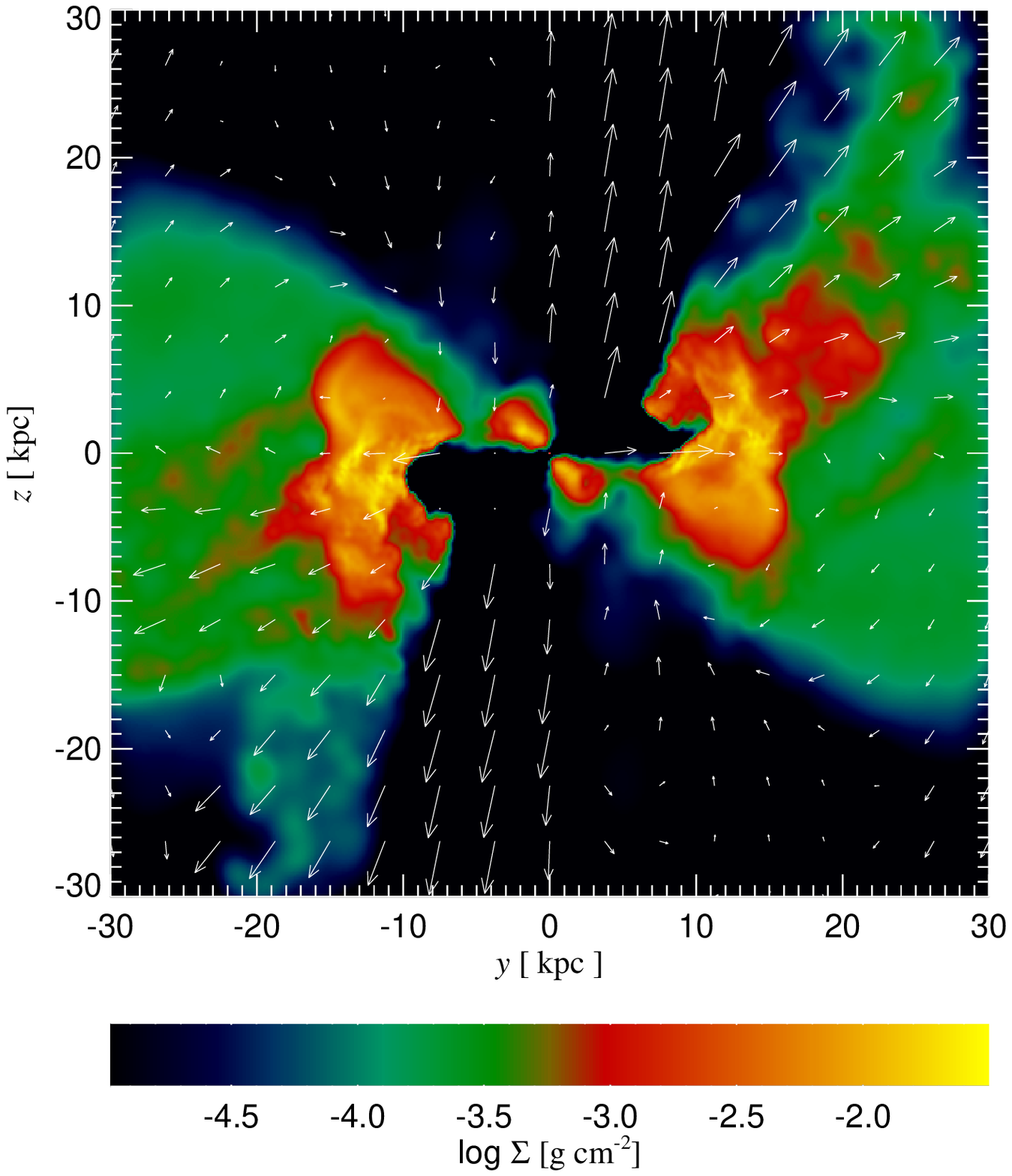,width=0.5\textwidth,angle=0}
\psfig{file=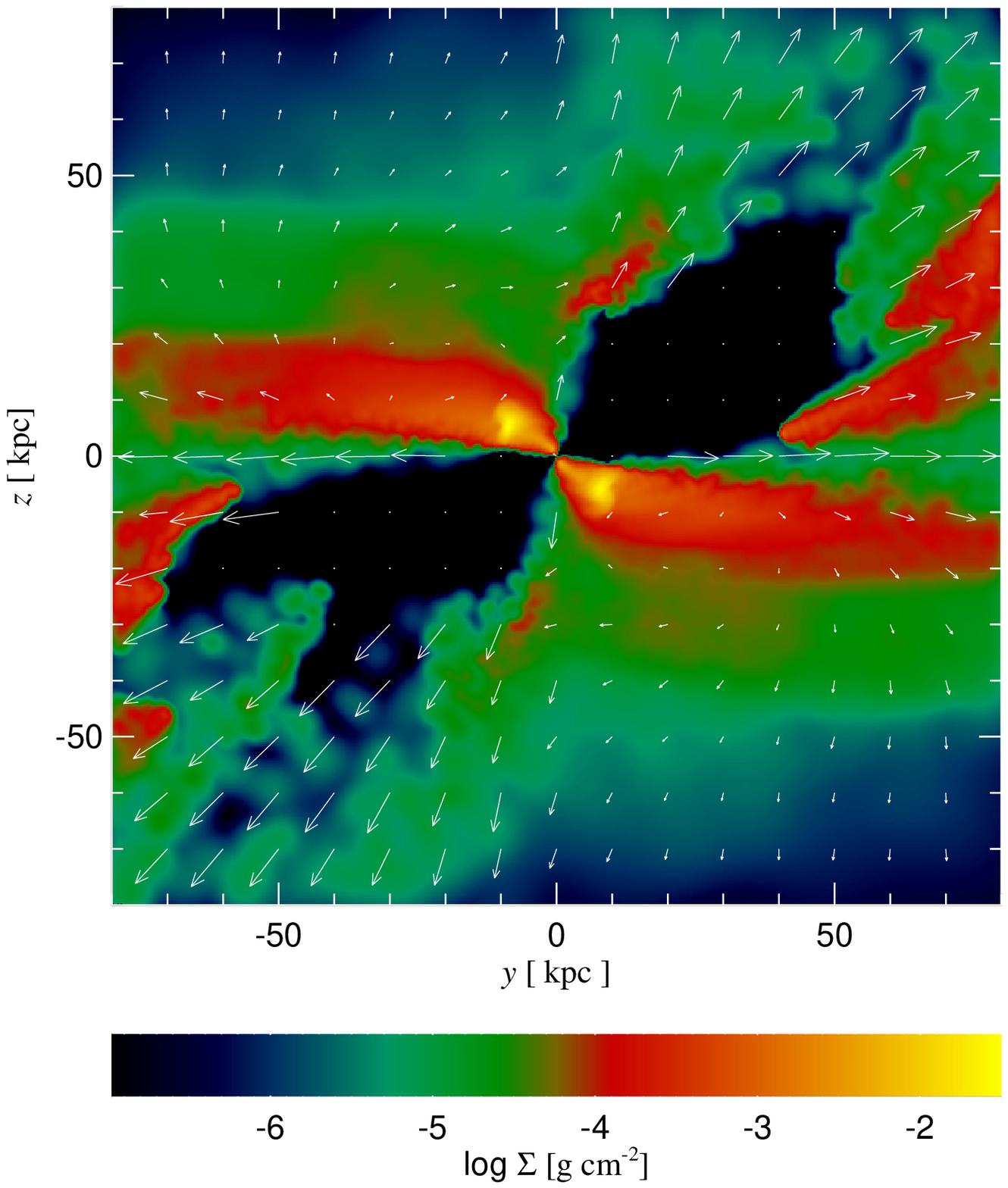,width=0.5\textwidth,angle=0}}
\caption{Angle-slice projected densities for the misaligned
  simulation at times $t=160$ (left panel) and $t=400$ (right panel)
  Myrs. The outflow eventually evacuates the directions along
  which it acts; only the inclined part of the disc that is shielded from the
  feedback survives to late times.}
\label{fig:fig1}
\end{figure}

Figure \ref{fig:fig1} shows projections of the gas surface density at times
$t=160$ Myrs (left panel) and $t=400$ Myrs (right panel).  While the gas
dynamics is very complex, one thing is clear: collimated feedback leads to the
separation of the gas into disc and outflow regions. While gas is driven off
to infinity along the axis of the outflow, perpendicular to the axis, where
there is no direct feedback, gas is able to accrete on the SMBH.

Summarising, more realistic models with rotation and/or anisotropic feedback
allow SMBHs to feed via a disc or regions not exposed to SMBH winds. In these
more realistic cases it is not clear why a robust $M_{\rm bh} - \sigma$
relation should be established.  In fact, some of the model predictions
contradict observations. For example, an isotropic SMBH wind impacting on a
disc (rather than a shell) of aspect ratio $H/R \ll 1$ requires the SMBH mass
to be larger by a factor $\sim R/H$, which is opposite to what is observed
\citep{Hu09}. In aspherical cases the SMBH outflows induce differential
motions in the bulge. This may pump turbulence that is known to hinder star
formation in star forming regions. SMBH feedback thus may not only drive gas
out of the bulge but also reduce the fraction of gas turned into stars.

\section{Conclusions}

A further progress in understanding of feedback requires careful, physics-based,
models of both accretion of gas and the effects that SMBH/NCs produce on their
environment through energy and momentum deposition.


\acknowledgements The author thanks Alex Hobbs and Andrew King for fruitful
discussions on interstellar gas turbulence and AGN feeding.




\end{document}